\documentclass{article}
\usepackage{spconf,amsmath,graphicx}
\usepackage{subcaption}
\usepackage{amssymb}

\title{Subjective Assessment of High Dynamic Range Videos Under Different Ambient Conditions}%
\twoauthors
 {Zaixi Shang\sthanks{This research was sponsored by a grant from Amazon.com, Inc., and by grant number 2019844 for the National Science Foundation AI Institute for Foundations of Machine Learning (IFML).
Zaixi Shang and Joshua Ebenezer contributed equally to this work.}, Joshua P. Ebenezer, Alan C. Bovik}
	{The University of Texas at Austin\\
	Department of Electrical and Computer Engineering\\
	 Austin, TX}
 {Yongjun Wu, Hai Wei, Sriram Sethuraman}
	{Amazon Prime Video\\
	Seattle, WA}
\begin{document}
%
\maketitle
\begin{abstract}
High Dynamic Range (HDR) videos can represent a much greater range of brightness and color than Standard Dynamic Range (SDR) videos and are rapidly becoming an industry standard. HDR videos have more challenging capture, transmission, and display requirements than legacy SDR videos. With their greater bit depth, advanced electro-optical transfer functions, and wider color gamuts, comes the need for video quality algorithms that are specifically designed to predict the quality of HDR videos. Towards this end, we present the first publicly released large-scale subjective study of HDR videos. We study the effect of distortions such as compression and aliasing on the quality of HDR videos. We also study the effect of ambient illumination on perceptual quality of HDR videos by conducting the study in both a dark lab environment and a brighter living-room environment. A total of 66 subjects participated in the study and more than 20,000 opinion scores were collected, which makes this the largest in-lab study of HDR video quality ever. We anticipate that the dataset will be a valuable resource for researchers to develop better models of perceptual quality for HDR videos.
\end{abstract}
\section{Introduction}

High Dynamic Range (HDR) Imaging is a set of techniques to extend the range of luminances and color that can be represented in an image or video. HDR10 is an open HDR standard that was announced by the Consumer Technology Association in 2015~\cite{cta} and is the most widespread of HDR formats. HDR10 content must have a bit-depth of 10 bits, the Rec. 2020~\cite{bt2020} color primaries (which cover 75.8\% of the CIE 1931 color space), the SMPTE ST 2084~\cite{smpte2084} Eletro-Optical Transfer Function (EOTF) (also known as the Perceptual Quantizer (PQ) EOTF), and contain static metadata regarding the color volume of the mastering display, the maximum frame-average light level, and the maximum content light level. HDR10 has seen increasing adoption in the past few years.
\par
Despite the efforts in VQA algorithms and databases \cite{9298463,9477407,9506189,9053634,8803179,9806466}, there remains challenges to predict the quality of user experience for HDR10 videos. The increase in the bit depth and the different transfer functions may change the ways in which distortions manifest and are perceived. Currently, no publicly available database of HDR10 content exists. Existing datasets of HDR content are either not publicly available or are based on obsolete standards.  

In the past few years a number of efforts have been made to create video quality datasets for HDR but they have all suffered from severe limitations, either due to the fast paced development of HDR standards that have made them obsolete or the inability of the authors to release the data publicly due to copyright issues. Azimi et al.~\cite{azimi2018evaluating} conducted a study with 18 subjects using 5 different 12-bit YUV contents captured from a RED Scarlet-X Camera and 5 different distortions, for a total of 30 videos. Pan et al.~\cite{pan2018hdr} conducted a study on the effect of compression on HDR quality with 6 source videos that were encoded using the PQ and HLG EOTFs and the BT. Baroncini et al.~\cite{baroncini2016verification} conducted a study with 12 videos and 40 subjects to evaluate the performance of HDR codecs. Rerabek et al.~\cite{rerabek2015subjective} conducted a study with 5 videos and 4 compression levels to evaluate objective quality assessment algorithms for HDR. The videos were limited to a resolution of 944$\times$1080 and the data was tone-mapped to an 8 bit format before playing. Athat et al.~\cite{athar2019perceptual} conducted a subjective study of HDR content that was HDR10 compliant and 14 source contents were compressed with H.264 and HEVC to generate 140 distorted videos that were watched by 51 subjects. Narwaria et al.~\cite{narwaria2015hdr} created a database with 10 HDR videos but they are not HDR10. Importantly, the data from \cite{azimi2018evaluating,pan2018hdr,baroncini2016verification,rerabek2015subjective,athar2019perceptual} were never made publically available for various issues.\par

We present the first open-sourced large-scale video quality database created for modern HDR10 videos. The database consists of 310 videos created from 31 reference contents that have been distorted by compression and aliasing. The videos were presented to subjects in two ambient conditions using dedicated hardware in a highly controlled environment. In contrast to these, our study is compliant with the most widely used modern HDR standard, containing all HDR, Wide Color Gamut (WCG) and High Frame Rate (HFR) videos. The dataset we have created is thus the largest and first publicly-available HDR10 video quality dataset. The new resources will be publicly available at http://live.ece.utexas.edu/research/Quality/index.htm.

\section{Details of subjective study}
\subsection{Contents}

\begin{figure*}
\centering
  \begin{subfigure}[t]{.16\linewidth}
    \centering\includegraphics[width=0.99\textwidth]{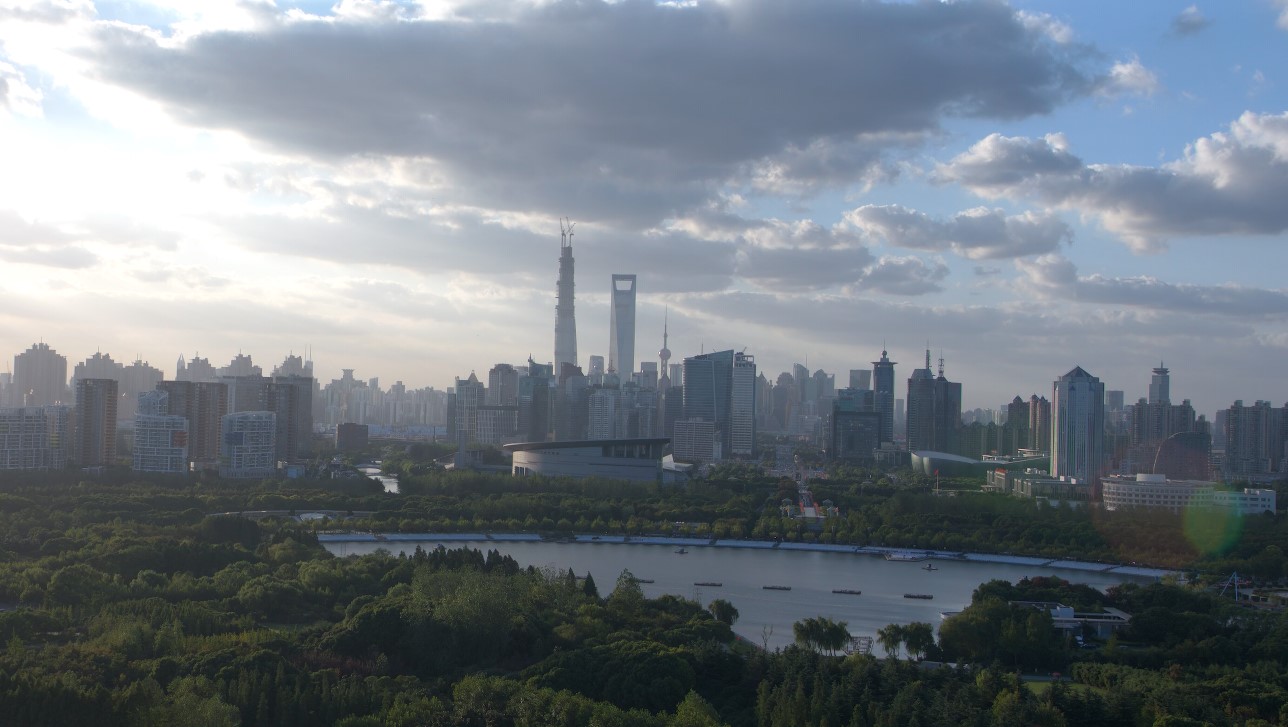}
  \end{subfigure}
  \begin{subfigure}[t]{.16\linewidth}
    \centering\includegraphics[width=0.99\textwidth]{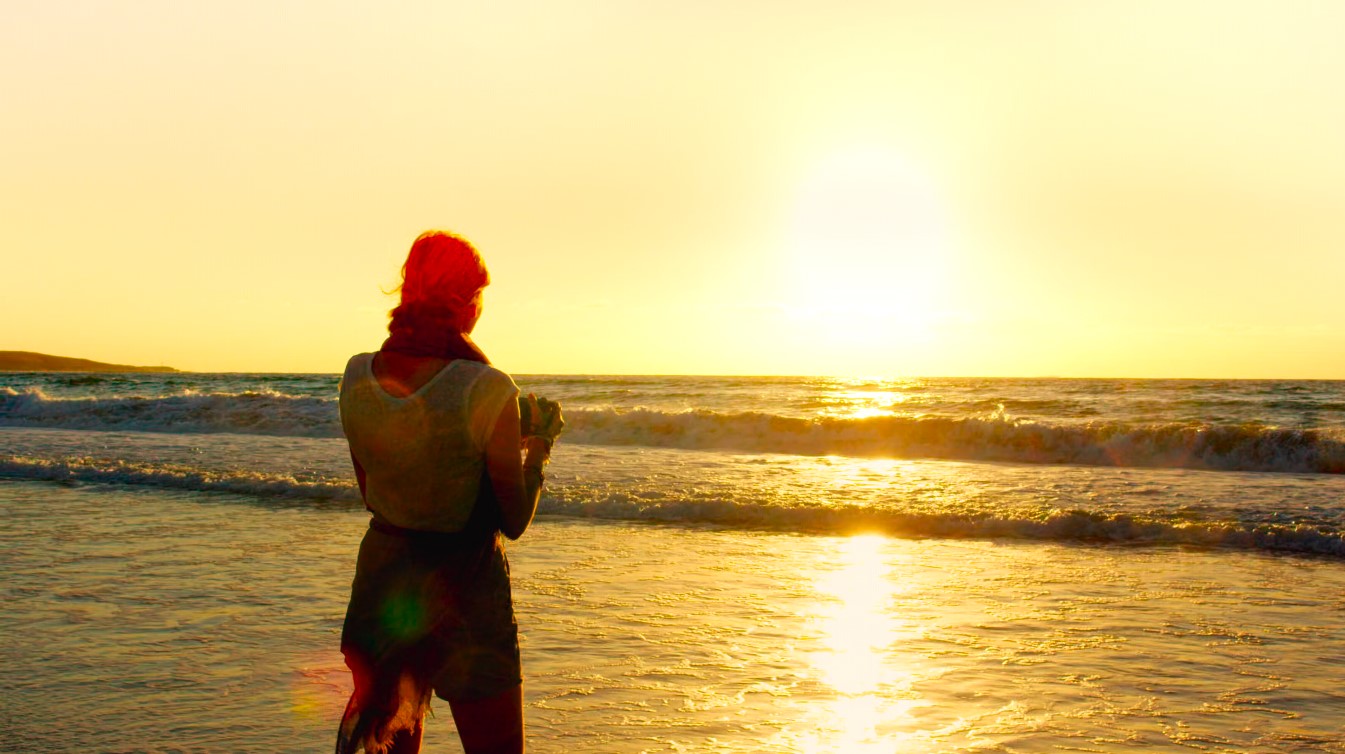}
  \end{subfigure}
  \begin{subfigure}[t]{.16\linewidth}
    \centering\includegraphics[width=0.99\textwidth]{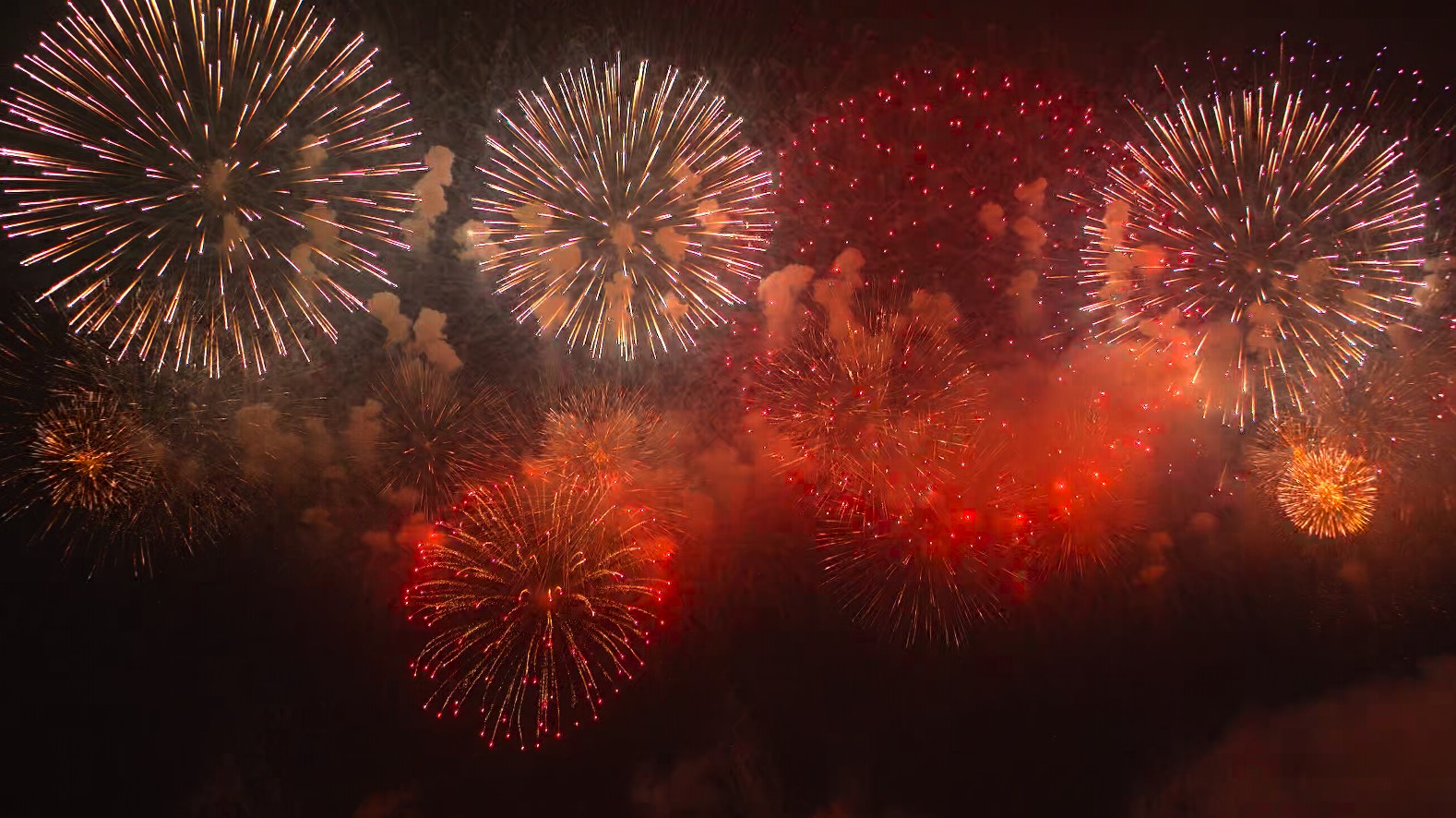}
  \end{subfigure}
   \begin{subfigure}[t]{.16\linewidth}
    \centering\includegraphics[width=0.99\textwidth]{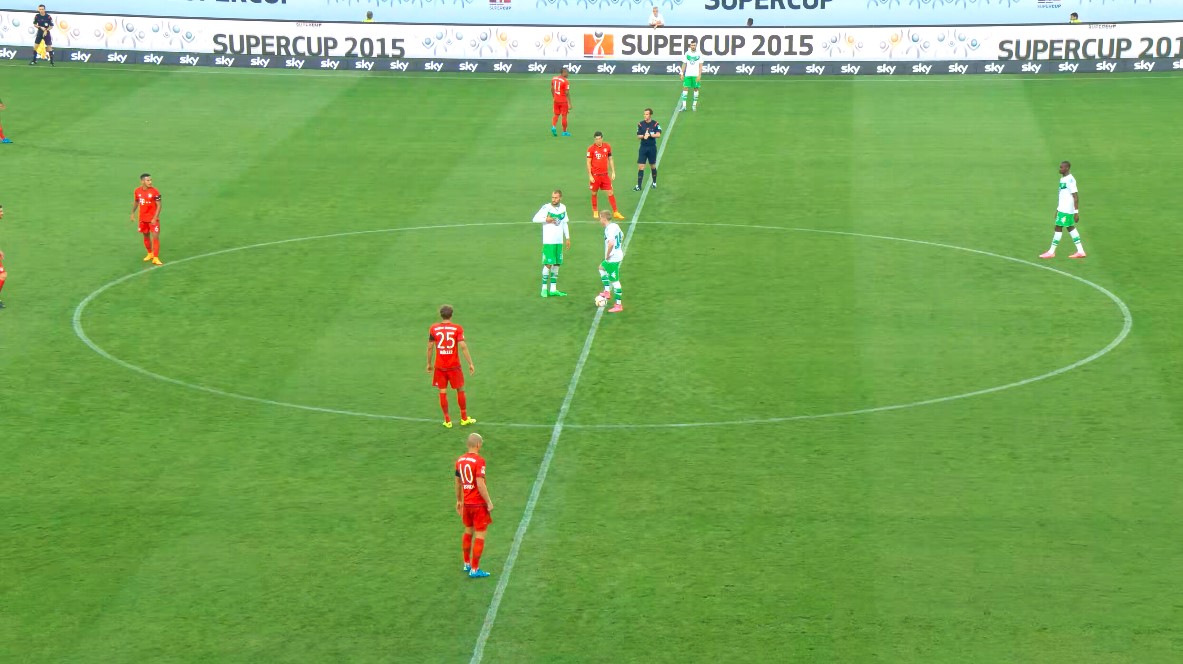}
  \end{subfigure}
  \begin{subfigure}[t]{.16\linewidth}
    \centering\includegraphics[width=0.99\textwidth]{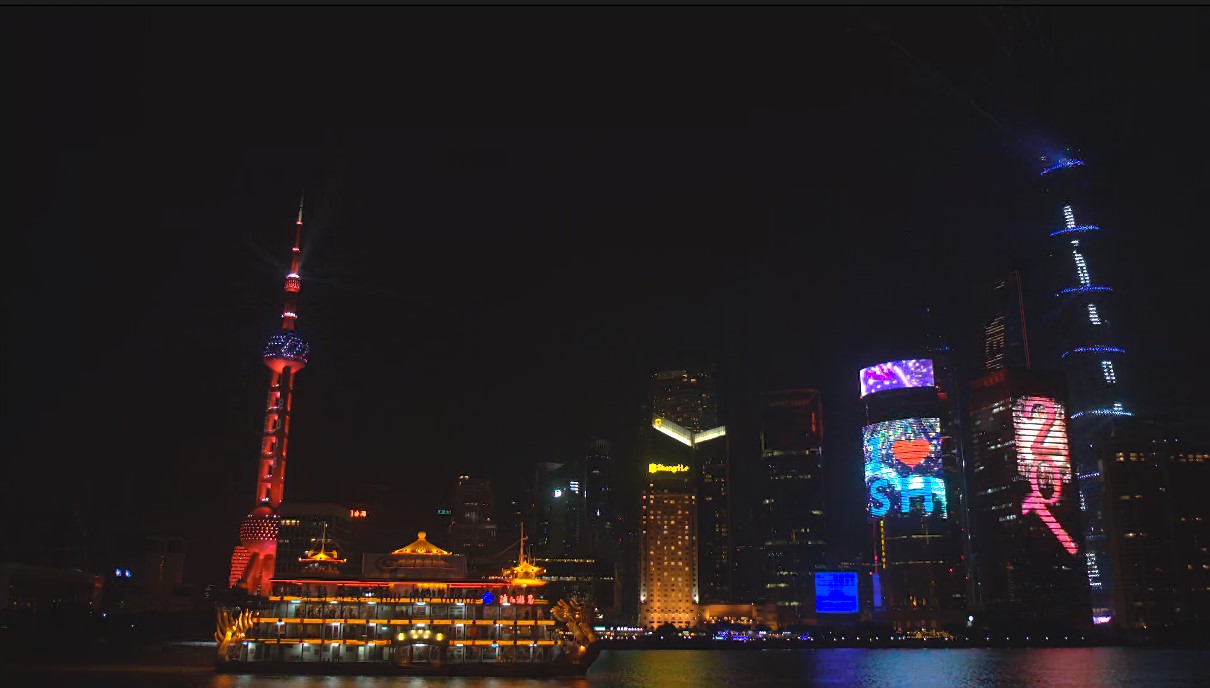}
  \end{subfigure}
  \begin{subfigure}[t]{.16\linewidth}
    \centering\includegraphics[width=0.99\textwidth]{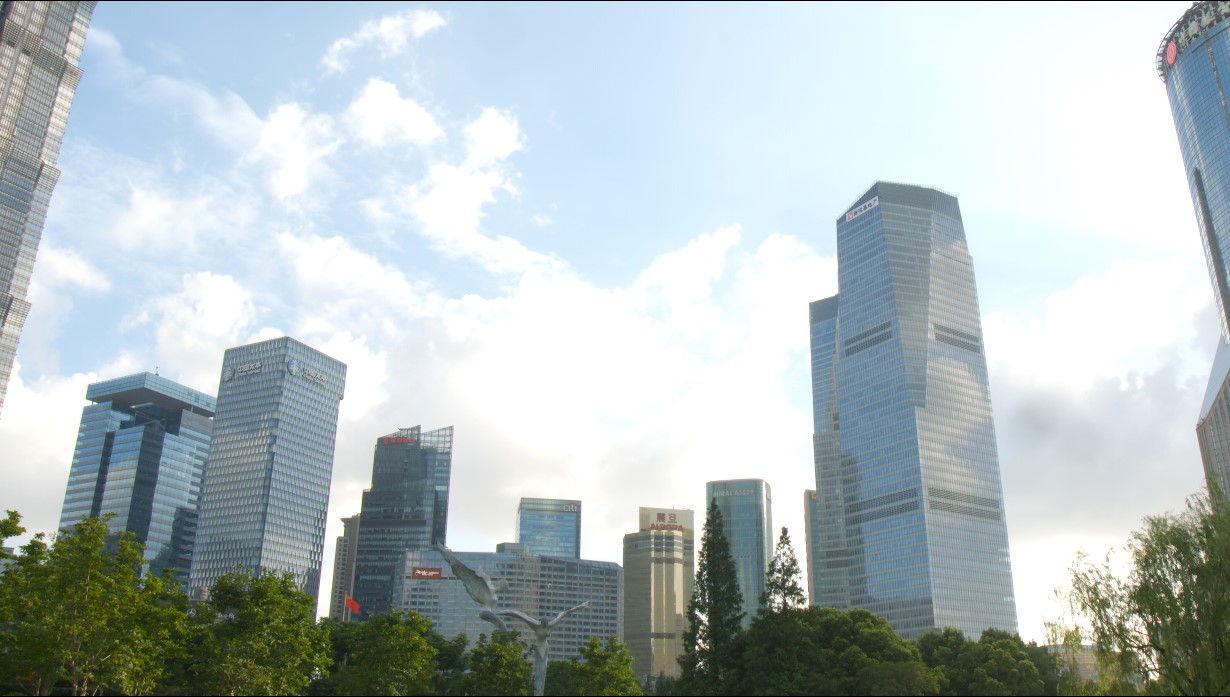}
  \end{subfigure}
\caption{\label{frames} Exemplar screenshots of frames from source sequences.}
\end{figure*}

We gathered high-quality, distortion-free source HDR10 sequences from a variety of web sources, including \cite{CDVL,SJTU,4kmedia}. These videos are captured with professional, high-end equipment. The source sequences all have a resolution of 3840x2160 pixels, frame rate of 50-60 fps and were progressively scanned with audio removed. All of the source sequences are HDR10 videos. We carefully cut the lengthier video sequences along the temporal dimension into one or more shorter clips of 7-10 seconds with no overlap to prevent the potential bias produced by different video lengths. The video durations vary to prevent awkward scene cuts in the middle of motion. Figure~\ref{frames} shows images of several sample frames from the source sequences we acquired, which span a wide range of scenes. 

\subsection{Test sequences}

Using the High Efficiency Video Coding (HEVC) Codec, we constructed 9 distorted video sequences from each source sequence, mixing various bitrates and resolutions. To reflect the HDR video streaming practice, we picked several bitrate and resolution combinations. The bitrate and resolution settings were chosen to guarantee that the distorted videos created are perceptually distinguishable and span a wide range of perceptual quality. The final bitrate and resolution settings are shown in Table~\ref{tab:levels}. We also included the source video sequences in the database as reference videos for the calculation of difference mean opinion scores (DMOS). The database has four spatial resolutions, and the 4K resolution and the 1080p resolution contain four and three bitrates respectively, mimicking the bitrate ladders used in HDR video streaming. The final video database contains 279 distorted videos and 31 reference videos, with a total of 310 videos.  

\begin{table}
\begin{center}
\begin{tabular}{|l|c|c|}
\hline
Number & resolution & bitrate (Mbps) \\
\hline\hline
1 & 3840×2160 & 15 \\
2 & 3840×2160 & 6\\
3 & 3840×2160 & 3\\
4 & 1920×1080 & 9\\
5 & 1920×1080 & 6\\
6 & 1920×1080 & 1\\
7 & 1280×720 & 4.6\\
8 & 1280×720 & 2.6\\
9 & 960×540 & 2.2\\
\hline
\end{tabular}
\end{center}
\caption{The bitrate and resolution settings for the distorted videos.}
\label{tab:levels}
\end{table}

\subsection{Subjective testing details}

The study was conducted at the University of Texas at Austin. A Samsung 65 inches Class Q90T QLED 4K UHD HDR Smart TV \cite{TV} was used to conduct the study. The TV was calibrated for HDR by an Imaging Science Foundation (ISF) certified professional using a Calman Calibration kit. The TV was connected to a workstation with a 12 GB Titan X Graphics Processing Unit (GPU) via a HDMI 2.0b cable which is capable of transferring videos at their original framerate for smooth playback of the videos. The workstation had a Windows 10 operating system with HDR enabled. The Potplayer Video Player was used for playback with the MadVR renderer. All advanced temporal processing options on the TV were disabled. \par 
The viewing distance was 1.5H, where H was the height of the TV. Subjects would watch each video and then see a screen where they were asked to present a quality score for the video that they had just seen. Subjects could then choose a quality score on a slider on the screen using their mouse. The slider was continuous and had five verbal markers at uniform intervals that said ``Bad", ``Poor", ``Fair", ``Good", and ``Excellent". The scores given by the subjects were sampled as integers from [0, 100] although numerical values were not made visible to the subjects.

 Two ambient conditions were used in this study in order to test the effect of ambient illumination on the perceived quality of HDR content. The first was a dark viewing condition that corresponded to a lab-study environment with an incident illumination on the TV of 5 lux, following the recommendation in \cite{bt2100} for critical viewing of HDR content and the recommendation in \cite{bt500} for general viewing conditions for a subjective study in a laboratory environment. A table lamb and a floor lamp were placed on either side of the TV to produce the light necessary for this environment. 
 \par
The second ambient condition was a living room environment. A pair of Neewer LED lights were used to produce an incident illumination on the TV of 200 lux for the living room environment, following the recommendation in \cite{bt500} for general viewing conditions for a subjective study in a home environment.

A total of 66 human subjects were recruited from the student population at the University of Texas at Austin. Each subject participated in two sessions separated by at least 24 hours, and viewed the videos in a randomized order. The subjects were divided into two groups, one for each ambient condition. 33 subjects watched the videos in the lab environment and 33 watched the videos in the living room environment. 
\section{Processing of Subjective Scores}

There are a number of ways in which subjective scores can be converted into Mean Opinion Scores (MOS). We compute the MOS as the average of subjective scores given by subjects (MOS), the average of $z$ scores (ZMOS), and also compute MOS using a statistical method proposed in ~\cite{li2020simple}.

\subsection{MOS}
Let $i_d$ be the index of subjects who viewed the videos in the dark lab environment and $i_b$ be the index of subjects who viewed the videos in the bright living-room environment. The MOS can be calculated as averages of the scores given by the subjects. 

\subsection{ZMOS}
The MOS for the videos can be calculated as the average of the $z$ scores as was done in \cite{pcs}. 
If the scores given by a subject $i_k$ for video $j$ are given by $s_{i_kj}$, the $z$ scores are given by 
\begin{equation}
    z_{i_kj} = \frac{s_{i_kj}-\mu_{i_k}}{\sigma_{i_k}}
\end{equation}
for $k=b,d$, subjects $i_d = 1,2 \dots S_d$, subjects $i_b = 1,2 \dots S_b$, and videos $j=1,2 \dots N$. For our database, $S_d=33, S_b =33$ and $N=310$. $\mu_{i_k}$ is the average score given by subject $i_k$ across all videos and $\sigma_{i_k}$ is the standard deviation of the scores given by subject $i_k$ across all videos.
Since there are two ambient conditions, there will be two sets of MOS. We refer to the MOS calculated from the $z$ scores as $ZMOS$, and we refer to the $ZMOS$ for the video $j$ whose scores were collected in the dark lab and bright living-room ambient conditions as $ZMOS_{dj}$ and $ZMOS_{bj}$, respectively.

\begin{equation}
    ZMOS_{kj} = \sum_{i_k=1}^{S_k} z_{i_kj}
\end{equation}
for $k=d,b$ and $j=1,2 \dots N$.  \par 

\subsection{SUREAL scores}
The deficiencies in the ITU BT 500.11 outlier removal method have been discussed and similar findings observed in \cite{li2020simple}. Therefore we used the method proposed in \cite{li2020simple}, called SUREAL, to find the Maximum Likelihood estimate of the scores. In this method, the opinion scores $s_{i_kj}$ are represented by a random variable $S_{i_kr}$
\begin{equation}
    S_{i_kj} = \psi_{kj} + \Delta_{i_k} + \nu_{i_k}X
\end{equation}
where $\psi_{kj}$ is the true quality of video $j$ in ambient condition $k$, $\Delta_{i_k}$ represents the bias of subject $i_k$, the non-negative term $\nu_{i_k}$ represents the inconsistency of subject $i_k$, and $X \thicksim N(0,1)$ are i.i.d. Gaussian random variables. The quantities $\psi_{kj}, \Delta_{i_k}, \nu_{i_k}$ are found by computing the log-likelihood of the observed scores and using the Newton-Raphson method to solve for the values of $\psi_{kj}, \Delta_{i_k}, \nu_{i_k}$ that maximize the log-likelihood. The method is robust to subject inconsistencies. 

\section{Effect of ambient illumination}

\begin{figure}[t]
     \centering
     \begin{subfigure}[b]{0.25\textwidth}
         \centering
         \includegraphics[width=\textwidth]{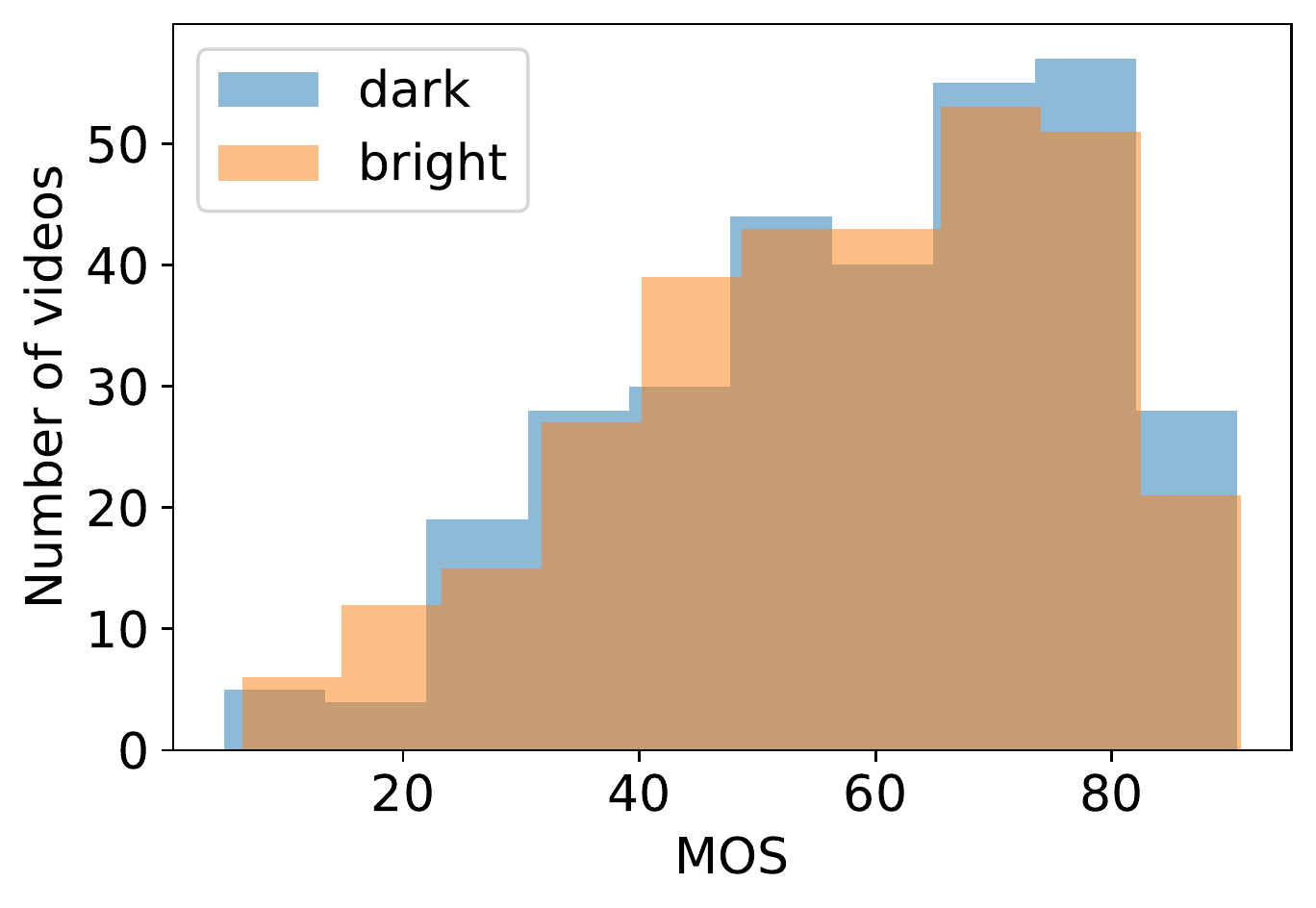}
         \caption{}
         \label{fig:hisa}
     \end{subfigure}
     \hfill
     \begin{subfigure}[b]{0.25\textwidth}
         \centering
         \includegraphics[width=\textwidth]{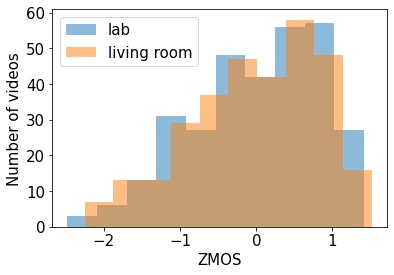}
         \caption{}
         \label{fig:hisb}
     \end{subfigure}
    \begin{subfigure}[b]{0.25\textwidth}
         \centering
         \includegraphics[width=\textwidth]{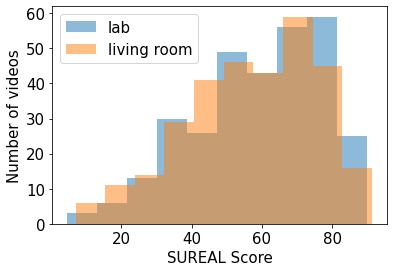}
         \caption{}
         \label{fig:hisc}
     \end{subfigure}
          \caption{Histograms showing the $MOS$, $ZMOS$ and SUREAL score distribution.}
        \label{fig:hist}
\end{figure}

\begin{figure}[t]
\begin{center}
   \includegraphics[width=0.98\linewidth]{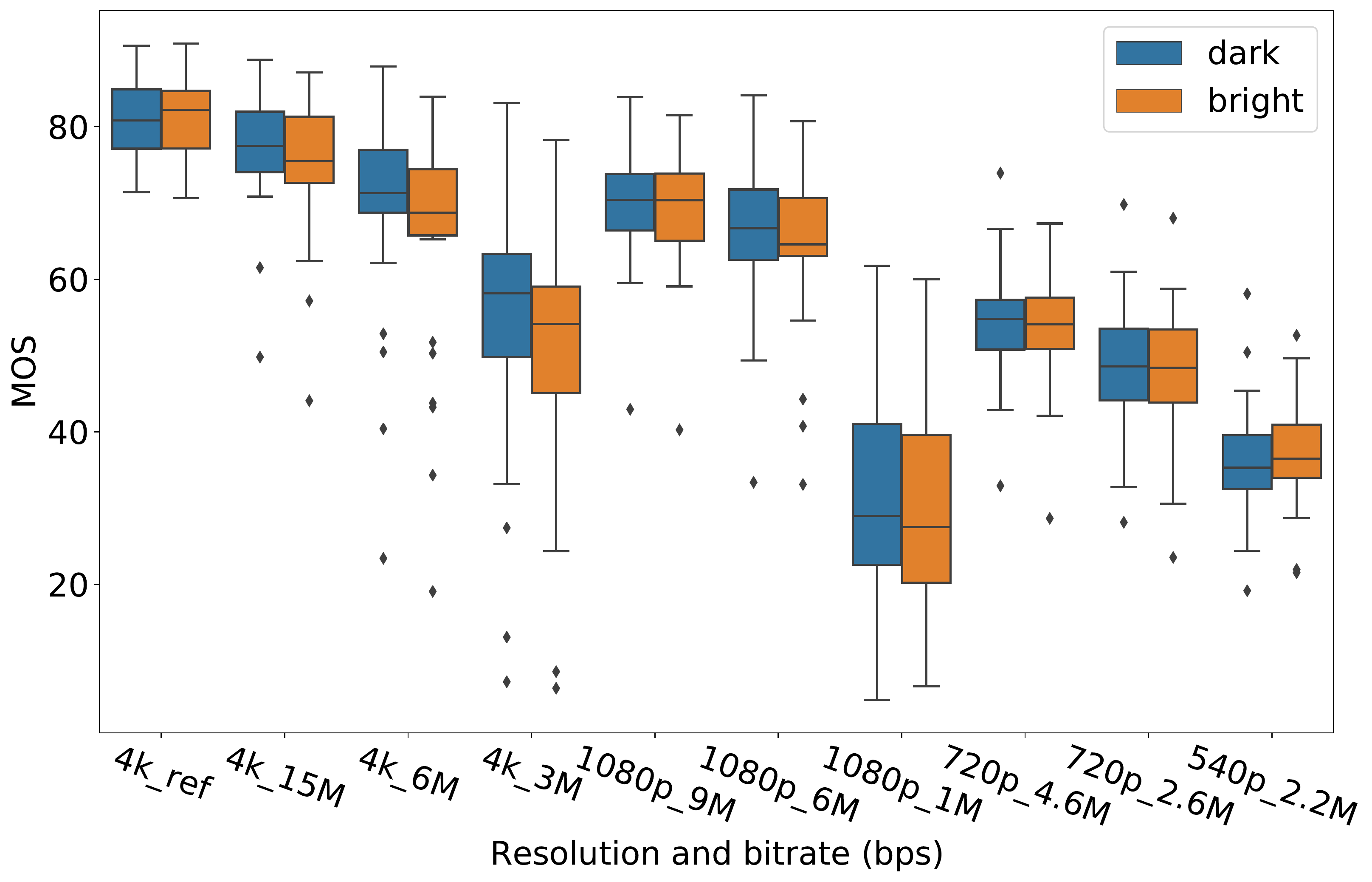}
\end{center}
   \caption{A box plot showing the distribution of MOS under two ambient illumination settings for each distortion combination.}
\label{fig:box}
\end{figure}

We used all three scores for the analysis of the effect of ambient illumination. It is worth noticing that $MOS$ and SUREAL scores preserve the differences between the absolute values of the perception scores in the two ambient conditions while $ZMOS$ scores don't because they are normalized. The distributions of the $MOS$, $ZMOS$ and SUREAL scores are shown in Figure~\ref{fig:hist}. The $MOS$ and SUREAL scores obtained in both ambient conditions cover a wide range of quality, and the overall distribution of scores in both ambient conditions are similar. Since SUREAL and MOS are absolute scores, one may deduce from Figure \ref{fig:hisa} and Figure \ref{fig:hisc} that the videos watched in the lab condition are rated slightly higher than those watched in the living room condition. The same conclusion cannot be drawn from Figure \ref{fig:hisb} because $ZMOS$ is a normalized score. Figure~\ref{fig:box} plots the $MOS$ against the spatial resolution and bitrate of all tested videos. The $MOS$ from both ambient illuminations covers a similar score range for each spatial resolution and bitrate combination, but the $MOS$ from the brighter living room setting has a slightly lower value than the one from the darker lab setting for most resolution and bitrate settings. This difference increases as the videos suffer from more severe distortion, i.e., lower bitrate and resolution.

To verify the differences that we observed in Figure~\ref{fig:box}, we conducted Welch's two-sided t-tests on the $MOS$ from both ambient illumination settings as well as the raw scores that we obtained from the study. We compared the $MOS$ of each resolution and bitrate settings, and the obtained p-value is shown in Table~\ref{tab:pvalue}. None of the resolution and bitrate combinations exhibit a p-value less than 0.05, indicating that although there exists a difference between the $MOS$ from different illumination settings, the difference between the mean $MOS$ for each combination is not statistically significant. We also compared the raw score obtained for each video. The result shows only 18 out of the 310 testing videos, have statistically significant differences between the average scores. There is not a strong pattern of the 18 videos, although 6 of the videos are very night scenes with large area covered under darkness, like firework and night traffic. 

\begin{table}
\begin{center}
\begin{tabular}{|l|c|c|c|}
\hline
Number & resolution & bitrate (Mbps) & p-value \\
\hline\hline
1 & 3840×2160 & ref & 0.5987 \\
2 & 3840×2160 & 15 & 0.1539 \\
3 & 3840×2160 & 6  & 0.1750 \\
4 & 3840×2160 & 3  & 0.1538\\
5 & 1920×1080 & 9  & 0.3422 \\
6 & 1920×1080 & 6  & 0.2856  \\
7 & 1920×1080 & 1  & 0.3105  \\
8 & 1280×720 & 4.6  & 0.4361\\
9 & 1280×720 & 2.6  & 0.3645 \\
10 & 960×540 & 2.2  & 0.7095 \\
\hline
\end{tabular}
\end{center}
\caption{The p-value of each resolution and bitrate.}
\label{tab:pvalue}
\end{table}

\par
We also used the confidence intervals from the SUREAL scores to consider the effect of ambient illumination. The SUREAL method provides 95\% confidence intervals for the scores using the Cramer-Rao bound. We found that for 10 videos, the confidence intervals did not overlap, indicating statistically significant differences. 7 out of these 10 videos had average luminances that were lower than the average luminance across all the videos but no other apparent relationship was found between these 10 videos. We also computed the 95\% confidence intervals for $MOS$ (assuming a normal distribution) and Only 4 videos had non-overlapping confidence intervals.

\section{Conclusion}

We conducted a large-scale subjective video quality study targeting HDR10 videos. The new resource includes 310 video sequences generated from 31 source contents using 10 combinations of bitrate and spatial resolution. We gathered and analyzed the subjective quality score that was gathered under two ambient illumination, and observed that the perceptual quality scores from the darker environment are slightly higher than the ones from the brighter environment. However, this difference is not statistically significant and it's not related to the average luminance of video sequences. The new database can be used to create, test, and compare both NR and FR video quality, assessment models. We are making the new database the first publicly available HDR10 VQA database.

\bibliographystyle{IEEEbib}
\bibliography{strings,refs}

\end{document}